\documentclass[draftcls, onecolumn, 10pt]{IEEEtran}
\usepackage{amssymb,amsmath}
\usepackage[dvips]{graphicx}
\usepackage{amsfonts}
\usepackage{subfigure}
\usepackage{booktabs,multirow}
\usepackage{color}
\usepackage{cite}
\usepackage[table]{xcolor}

\IEEEoverridecommandlockouts

\begin{document}

\title{Energy Efficiency Challenges of 5G Small Cell Networks}

\author{\normalsize
Xiaohu Ge$^1$,~\IEEEmembership{Senior~Member,~IEEE,} Jing Yang$^1$, Hamid Gharavi$^2$,~\IEEEmembership{Fellow,~IEEE}, Yang Sun$^1$\\
\vspace{0.70cm}
\small{
$^1$School of Electronic Information and Communications,\\
Huazhong University of Science and Technology, Wuhan 430074, Hubei, P. R. China.\\
%Email: \{xhge \}@mail.hust.edu.cn\\
Email: \{xhge, yang\_jing, tylzzq\}@mail.hust.edu.cn\\
\vspace{0.2cm}
$^2$National Institute of Standards and Technology
(NIST),\\
Gaithersburg, MD 20899-8920 USA.\\
Email: hamid.gharavi@nist.gov}\\
\thanks{\small{ Submitted to Green Communications and Computing Networks Series of IEEE Communications Magazine.}}
}

\renewcommand{\baselinestretch}{1.2}
\thispagestyle{empty}
\maketitle
\thispagestyle{empty}
\setcounter{page}{1}\begin{abstract}
The deployment of a large number of small cells poses new challenges to energy efficiency, which has often been ignored in fifth generation (5G) cellular networks. While massive multiple-input multiple outputs (MIMO) will reduce the transmission power at the expense of higher computational cost, the question remains as to which computation or transmission power is more important in the energy efficiency of 5G small cell networks. Thus, the main objective in this paper is to investigate the computation power based on the Landauer principle. Simulation results reveal that more than 50\% of the energy is consumed by the computation power at 5G small cell BS's. Moreover, the computation power of 5G small cell BS can approach 800 watt when the massive MIMO (e.g., 128 antennas) is deployed to transmit high volume traffic.  This clearly indicates that computation power optimization can play a major role in the energy efficiency of small cell networks.

\end{abstract}

\IEEEpeerreviewmaketitle

\newpage
\section{Introduction}
With the anticipated high traffic, small cell networks are emerging as an inevitable solution for 5G cellular networks \cite{1Ge}. In particular, the massive multiple input multiple-output (MIMO) and millimeter wave technologies are expected to be deployed towards improving the transmission rate and reduce the transmission power of 5G mobile communication systems \cite{2Andrews}. On the other hand, more computation power will be required to process anticipated heavy traffic at small cell base stations (BS's). Under these conditions, a tradeoff between computation and transmission power needs to be thoroughly evaluated in order to achieve energy efficiency optimization for 5G small cell networks.

This has been widely investigated in \cite{3Xge,4Samarakoon,5Ta,6Liu}. Compared with transmission power, computation power was obviously smaller and usually fixed as a constant in a traditional energy efficiency evaluation of BS's \cite{4Samarakoon}. As a consequence, the energy efficiency investigation of small cell networks has focused on the optimization of transmission power at BS's \cite{5Ta}. Furthermore, the BS sleeping scheme has been considered to improve energy efficiency where the radio frequency (RF) chains and transmitters of BS's are closed to save transmission power \cite{6Liu}.  In addition, the computation power of small cell BS's has been improved by the volume and complexity of signal processing, which is weighted by massive MIMO and millimeter wave technologies \cite{7Choi}.

When small cell BS's are ultra-densely deployed in 5G cellular networks\cite{8Chen}, there exist scenarios in which the computation power of BS's will become larger than the transmission power of BS's despite lower power transmission requirements for small cell BSs.

The transmission rate of 5G mobile communication systems is expected to reach to an average of 1 Gbps (10 Gbps at the peak rate) \cite{2Andrews}. Hence, the huge traffic has to be handled at the base band units (BBU's) of small cell BS's and then the computation power of signal processing has to be accordingly improved at BBU's. Moreover, the cache communications and cloud computing network architecture will strengthen functions of signal processing and computing at small cell BS's. Nonetheless, the computation power of 5G small cell networks could be predicted to increase in the near future. All the above reasons trigger us to rethink the roles of computation and transmission power in 5G small cell networks.

Based on the Landauer principle, we first proposed a computation power model for 5G small cell networks. Considering that the massive MIMO and millimeter wave technologies are adopted at small cell BS's, the impact of the number of antennas and bandwidths on the computation power of 5G small cell networks is investigated. Simulation results indicate that the computation power will consume more than 50\% of the energy at 5G small cell BS's. It is a surprising result for the energy efficiency optimization of 5G small cell networks. Finally, future challenges of energy efficiency optimization are discussed for 5G small cell networks and conclusions are drawn in the last section.

\section{Power Consumption at BSs}
To evaluate roles of computation and transmission power for BS's, the total BS power consumption needs to be analyzed in detail. Therefore, in this section 5G transmission technologies, such as massive MIMO and millimeter wave technologies will be incorporated for analyzing the power consumption of small cell BS's.

\subsection{BS Power Consumption Types}
Considering functions and architectures of BS's, the power consumption at BS's is typically classified into three types: transmission power, computation power and additional power which are described as follows.

\begin{itemize}
  \item
     The transmission power corresponds to the energy used by power amplifiers (PAs) and RF chains, which perform the wireless signals change, i.e., signal transforming between the base band signals and the wireless radio signals. Besides, the power consuming at feeders is included as a part of the transmission power.
  \item
     The computation power represents the energy consumed at base band units (BBU's) which includes digital single processing functions, management and control functions for BS's and the communication functions among the core network and BS's. All these operations are executed by softwares and realized at semiconductor chips.
  \item
      The additional power represents the BS power, except for the transmission and computation power, e.g., the power consumed for maintaining the operation of BS's. More specifically, the additional power includes the power lost at the exchange from the power grid to the main supply, at the exchange between different direct-current to direct-current (DC-DC) power supply, and the power consumed for active cooling at BS's.
\end{itemize}

The values of three types of consumed power are different depending on the types of BS. For example, unlike the macro cell BS the small cell BS normally does not have the active cooling system.

\subsection{Total BS Power Consumption Model}
The EARTH project has promoted energy efficiency optimization for wireless access networks and proposes a framework for the power consumption at BS's\cite{9Imran}. Based on this energy efficiency framework, the BS is divided into five parts (see Fig. 1): the antenna interface, the power amplifier, the RF chains, the BBU, the mains supply, and cooling and direct-current to direct-current (DC-DC). The power consumed at the power amplifier and the antenna interface occupy the largest proportion of the total power consumption in macro cell BS's, i.e., 57\%. The portion of RF chains and BBU¡¯s is about 10\% and 13\%, respectively. And the proportion of remaining parts is about 20\%. To analyze this in greater detail, the total BS power consumption model is presented as follows: When the BS is equipped with ${N_{TRX}}$ antennas, the total BS power consumption ${P_{in}}$ is calculated by ${P_{in}} = \frac{{{P_{PA}} \cdot {N_{TRX}} + {P_{RF}} \cdot {N_{TRX}} + {P_{BB}}}}{{(1 - {\sigma _{DC}})(1 - {\sigma _{MS}})(1 - {\sigma _{cool}})}}$, where ${P_{PA}}$ is the power of PA per antenna, ${P_{RF}}$  is the RF chain power per antenna, ${P_{BB}}$ is the power consumed at the BBU, ${\sigma _{DC}}$ is the power loss rate of the DC-DC converter, ${\sigma _{MS}}$ is the power loss rate of the alternating current supply, and ${\sigma _{cool}}$ is the power loss rate of cooling. Based on the expression of the total BS power consumption, ${P_{PA}} \cdot {N_{TRX}} + {P_{RF}} \cdot {N_{TRX}}$ is the transmission power, ${P_{BB}}$ is the computation power, $(1 - {\sigma _{DC}})(1 - {\sigma _{MS}})(1 - {\sigma _{cool}})$ is the relationship between the power loss rate and the total power consumption. The power of PA per antenna is calculated by ${P_{PA}} = \frac{{{P_{out}}}}{{{\eta _{PA}} \cdot (1 - {\sigma _{feed}})}}$, where ${P_{out}}$ is the transmission power at every antenna, ${\eta _{PA}}$ is the exchange efficiency of PA, and feeder loss is configured as ${\sigma _{feed}}=-3$ dB. For macro cell BS's and small cell BS's, the values of ${\sigma _{DC}}$, ${\sigma _{MS}}$ and ${\sigma _{cool}}$ are configured as as 6\%, 7\%, 9\% and 8\%, 10\%, 0\%, respectively\cite{9Imran}. To simplify calculation, in this paper the RF chain power per antenna is usually fixed as different constants corresponding to different types of BS's. Since ${P_{BB}}$ is obviously less than the power consumed at other parts of BS's, the power consumed at the BBU is fixed as constant in a traditional BS power model\cite{10Desset}. Bear in mind that, as small cells are expected to be widely deployed in 5G cellular networks, the distance between BS's and users will be much shorter, resulting in a considerable reduction of transmission power. Under these conditions, the BBU becomes the dominant source of power consumption.

\section{Computation Power Model}
Because of the extensive traffic processing at 5G small cell BS's, the volume of data processing at 5G small cell BS's is evaluated by the operation per second at BBU's. Furthermore, Landauer's principle is used to estimate the computation power consumed for data processing in this section. In this section we also study the impact of massive MIMO and millimeter wave technologies on the computation power of 5G small cell BS's.

\subsection{Computation Power Types}
In traditional macro cell BS¡¯s the power used at BBU¡¯s (BBU is the core unit of a BS) is small compared with the power consumed by PA¡¯s. With the recent advances of 5G of the massive MIMO and millimeter wave technologies, small cell BS's are replacing macro cell BS's to perform the function of wireless data transmission in 5G cellular networks. Moreover, the power consumed at BBU's is expected to gradually increase because of the massive traffic in 5G small cell BS's.

Fig. 1 is a typical logistical architecture of eNodeB BS, i.e., a macro cell BS in a cellular network. Without a loss of generality, the BBU of a macro cell BS includes four systems: the base band system, the control system, the transfer system, and the power system. The detailed functions of these systems in BBU are described as follows.

\begin{itemize}
  \item
      The functions of a base band system include signal filtering, fast Fourier transform/inverse fast Fourier transform (FFT/IFFT), modulation and demodulation, digital-pre-distortion (DPD) processing, signal detection, and wireless channel coding/decoding. Note that, the function of signal processing used for transmitters and receivers is performed by the BBU.
  \item
      The control system takes charge of controlling and managing resource allocation at BS's in order to provide control interface between the BS and other network units. Moreover, communication control protocols are run at the control system. The control system also provide an interface of man-machine language (MML) for the local maintain terminal (LMT) to configure the resource allocation of BBU's.
  \item
      The transfer system connects with the mobility management entity/serving-gateway (MME/S-GW) of the core network by the S1 interface(see Fig. 1). Moreover, the control and management information among BS's are forwarded by the X2 interface of the transfer system in BBU's.
  \item
     The power system is responsible for power supply, cooling, and monitoring at BBU's.
\end{itemize}

For small cell BS's, most functions are integrated into a few semiconductor chips and there is not a single power system. Therefore, the systems of BBU's at small cell BS's is simpler than the systems of BBU's at macro cell BS's.

\subsection{Computation Power Model}
Based on the four systems in the logistical architecture shown in Fig.1, the main difficulty is how to calculate the computation power for every logistical system in BBU's. To achieve this, we partition a BBU into different parts based on the hardware architecture as shown in Fig. 2. These consist of DPD, Filter, CPRI, OFDM, FD, FEC, and CPU where DPD is the digital-pre-distortion processing part, filter is the hardware used for up/down signal sampling and filtering, CPRI is the common public radio interface part for connecting to the core network and RF chains by serial links, OFDM is the hardware used for FFT and orthogonal frequency-division multiplexing (OFDM)-specific signal processing, FD is the frequency-Domain processing part, which includes, symbol mapping/demapping and MIMO equalization, FEC is the forward error correction which includes the channel coding and decoding, and CPU is the BBU platform control processor. Based on Landauer's principle, we estimate the computation power of semiconductor chips using Giga operations per second (GOPS) and considering different semiconductor chip techniques. The computation power of BBU is summed up by the computation power of every hardware part, i.e., every semiconductor chip at BBU.

Landauer's principle was proposed in 1961 by Rolf Landauer who attempted to apply the thermodynamic theory to digital computers. Landauer's principle elaborates the relationship between the information process and energy consumption from the viewpoint of a microscopic degree of freedom in statistical physics. This is based on a physical principle pertaining to the lower theoretical limit of energy consumption that corresponds to the computation. Bear in mind that the concept of entropy in information theory introduced by Claude Shannon is borrowed from the thermodynamic theory. Similarly, Landauer's principle connects these two concepts of information and energy by using the thermodynamic theory and statistical physics. Therefore, in this paper Landauer's principle is first used to analyze the computation power consumption in 5G small cell networks. More specifically Landauer's principle points out that any logically irreversible manipulation of information, such as the erasure of a bit or the merging of two computation paths, must be accompanied by a corresponding entropy increase in non-information-bearing degrees of freedom of the information-processing apparatus or its environment\cite{11Landauer}. In other words, erasing a bit information will consume more than $kT\ln (2)$ energy in a computing system, where $k$ is the Boltzmann constant, i.e., $1.38 \times {10^{ - 23}}Joule/Kelvin$, $T$ is the kelvin temperature\cite{12Berut}. According to Landauer's principle, the lower bound of computation power for a computing system can be obtained. Compared with the value of computation power at real semiconductor chips, there exists a difference of three orders of magnitude for the values of computation power derived by Landauer's principle \cite{13Cockshott}. Moreover, the values of computation power are different when different semiconductor chip techniques are adopted at BBU's. Under these conditions, the main difficulty is how to accurately calculate the computation power of small cell BS's using the Landauer's principle.

To overcome the gap of computation power estimated by Landauer's principle and real semiconductor chips, we propose a power coefficient $\varepsilon $ is that can represent the level of the semiconduct chip technique in BBU's. Moreover, the power coefficient $\varepsilon $ is defined as the ratio of the active switching power of a transistor and the limit of Landauer's principle. From Fig. 3, the power coefficient $\varepsilon $ reflects the distance between semiconductor chip techniques and the limit of Landauer's principle. Bear in mind that up till now the development of semiconductor chip techniques still follows Moore's law. However, the international technology roadmap for semiconductors (ITRS) predicts that the development of semiconductor chip techniques will deviate from Moore's law when the power coefficient approaches the limit of Landauer's principle. For example, when nanomagnetic Logic is used for chips, the computation power is expected to approach the limit of Landauer's principle \cite{14Lambson}. Considering the development of current chip techniques, we focus our attention on the computation power of semiconductor chips.

Without a loss of generality, in this paper the power coefficient is configured as $\varepsilon = {10^3}$ when the 22 nanometer semiconductor technique is assumed to be adopted for chip manufacture in BBU's. Moreover, the active switching power of a transistor is approximated by ${E_{FET}} \approx \varepsilon kT\ln (2)$, which is used to calculate the power for operating 1 bit information at the semiconductor chip of BBU's.

In general, the data processing rate of semiconductor chips is represented by the instructions per second (IPS). Based on the definition of GOPS, in this paper the relationship between the IPS and the GOPS is expressed by $IPS = \frac{{GOPS \times {{10}^9}}}{{64}}$ when the logistical architecture of semiconductor chips is assumed to be 64 bit. According to the experimental  results in \cite{15Zhirnov}, the information throughput of semiconductor chips is denoted by $\rho  = {(\frac{{IPS}}{\omega })^{\frac{1}{\gamma }}}$, where $\omega$ and $\gamma$ are configured as 0.1 and 0.64, respectively. As a consequence, the computation power of different parts of a BBU is calculated by the product of the information throughput of semiconductor chips and the active switching power of transistors considering different values of GOPS at different parts of the BBU.

Since different types of BS's have different hardware components at BBU's, it is difficulty to directly build a uniform model to evaluate the computation power of BBU's in different types of BS's. Therefore, we first build a reference BS with typical parameters. By comparing different types of BS's with reference BS, we can derive the computation power of different BBU's for different types of BS's. Without a loss of generality, the system parameters are represented by $i \in \{ BW,Ant,M,R,dt,df\}$, where $BW$ is the bandwidth parameter, $Ant$ is the number of antennas parameter, $M$ is the modulation coefficient parameter, $R$ is the parameter of coding rate, $dt$ is the parameter of time-domain duty-cycling, and $df$ is the parameter of frequency-domain duty-cycling. To simplify symbols in this paper, $X_i^{ref}$ is denoted as the reference BS. When the subscript $i$ of $X_i^{ref}$ is replaced by different symbols, the new variable represents the corresponding system parameter in the reference BS, e.g., $X_{BW}^{ref}$ is the bandwidth of the reference BS. Similarly, $X_i^{real}$ is denoted for a real BS. When the subscript $i$ of $X_i^{real}$ is replaced by different symbols, the new variable represents the corresponding system parameter in the real BS, e.g., $X_{BW}^{real}$ is the bandwidth of the real BS.

Considering different computation powers at different hardware parts of a BBU, the computation power of DPD (Digital Pre-Distortion), Filter, CPRI (Common Public Radio Interface), OFDM, FD (Frequency-Domain), FEC (Forward Error Correction) and CPU at reference BS are denoted by $P_{DPD}^{ref}$, $P_{Filter}^{ref}$, $P_{CPRI}^{ref}$, $P_{OFDM}^{ref}$, $P_{FD}^{ref}$, $P_{FEC}^{ref}$ and $P_{CPU}^{ref}$, respectively. The different hardware parts of a BBU depend on the different system parameters of BS. ${S_i}$, $i \in \{ BW,Ant,M,R,dt,df\}$ signifies the ratio of the different hardware parts of the BBU and the system parameters of the BS. When the relationship between the hardware part of BBU and the system parameter of BS is linear, the corresponding ${S_i}$ is configured as 1. If such a relationship is non-linear, the corresponding ${S_i}$ is set to 2. When the relationship between the hardware part of BBU and the system parameter of BS is independent, the corresponding ${S_i}$ is configured as 0. The detailed configuration parameters of ${S_i}$ are illustrated in Table I.

Based on measurement results from the reference BS, the computation power of the reference BBU can be obtained by $P_{BB}^{ref} = P_{DPD}^{ref} + P_{Filter}^{ref} + P_{CPRI}^{ref} + P_{OFDM}^{ref} + P_{FD}^{ref} + P_{FEC}^{ref} + P_{CPU}^{ref}$. To calculate the computation power of real BBU's, the reference coefficient $\alpha$ is defined by $\alpha  = \prod\limits_i {{{\left( {\frac{{X_i^{real}}}{{X_i^{ref}}}} \right)}^{{S_i}}}}  = {\left( {\frac{{X_{BW}^{real}}}{{X_{BW}^{ref}}}} \right)^{{S_{BW}}}} \cdot {\left( {\frac{{X_{Ant}^{real}}}{{X_{Ant}^{ref}}}} \right)^{{S_{Ant}}}} \cdot {\left( {\frac{{X_M^{real}}}{{X_M^{ref}}}} \right)^{{S_M}}} \cdot {\left( {\frac{{X_R^{real}}}{{X_R^{ref}}}} \right)^{{S_R}}} \cdot {\left( {\frac{{X_{dt}^{real}}}{{X_{dt}^{ref}}}} \right)^{{S_{dt}}}} \cdot {\left( {\frac{{X_{df}^{real}}}{{X_{df}^{ref}}}} \right)^{{S_{df}}}}$. Finally, the computation power of real BBU's is calculated by $P_{BB}^{real} = \alpha  \cdot P_{BB}^{ref}$.

\section{Evaluations of Computation Power}

Considering that 5G small cell networks with massive MIMO's and millimeter wave techniques have not yet been commercially deployed, it is difficult to compare our simulation results with real 5G small cell networks. To validate the performance of the proposed power consumption model, we first compare the results of the proposed model with those of the EARTH project \cite{9Imran}, which measures the power consumption of macro cell and small cell BS's from real wireless networks. Without a loss of generality, the two wireless communication systems are configured with 10 MHz and 2$\times $2 antennas at BSs and terminals. Based on the results from the EARTH project, the total power consumption and computation power of a macro cell BS are 321.6 W and 29.68 W, respectively. Similarly, for our proposed power consumption model the total power consumption and the computation power of a macro cell BS are 317.84 W and 24.78 W, respectively. In the case of a small cell BSS for the EARTH project, the total power consumption and computation power of a small cell BS are 6.2 W and 2.4 W, respectively. For the proposed power consumption model, the total power consumption and computation power are 7.22 W and 3.6 W, respectively. Compared with the above, the results of the proposed power consumption model are in agreement with the results of real wireless networks. Therefore, our proposed power consumption model is shown to be capable of estimating the power consumption of 5G small cell networks.

Without loss of generality, the system parameters of the reference BS are configured as $X_{BW}^{ref} = 20$ MHz, $X_{Ant}^{ref} = 1$, $X_M^{ref} = 6$, $X_R^{ref} = 1$, $X_{dt}^{ref} = 100\%$, $X_{df}^{ref} = 100\%$. Based on the configuration parameters of the BS's in Table I, the computation power of BS's is simulated for 5G small cell networks. Since the massive MIMO and millimeter wave technologies are the core technologies for 5G mobile communication systems, in this section the impact of the number of antennas and bandwidths on macro cell BS's and small cell BS's are simulated in detail.

Generally speaking, the PA's of macro cell BS and small cell BS are configured as 102.6 W and 1.0 W. Fig. 4 illustrates the computation power of BS with respect to the number of antennas and bandwidths. The default system parameters of real BS's are configured as follows: the bandwidth is 20 MHz, the modulation is 64-quadrature amplitude modulation (QAM), the coding rate is $\frac{5}{6}$, the time-domain duty-cycling is 100\%  and the frequency-domain duty-cycling is 100\%. Fig. 4(a) shows the computation power of BS's with respect to the number of antennas. Based on the results in Fig. 4(a), the computation power of BS's quickly increases with the increase the number of antennas. The reason is that the computation power consumed for frequency-Domain processing is in proportion to the square of the number of antennas. Moreover, the computation power of macro cell BS's is always larger than the computation power of small cell BS's when the number of antennas is increased. When the number of antennas is equal to 128, i.e., adopting the massive MIMO technology, the computation power of macro cell BS is larger than 3000 W and the computation power of small cell BS is larger than 800 W.

In general, with the adaptation of millimeter wave techniques, 5G communication systems will be able to support large bandwidths (e.g., 400 MHz), or more precisely, high transmission rates. Consequently, this would require more processing at the BBU, hence further increasing the computation power at BS's. Therefore, in this paper, the impact of a millimeter wave technique on the computation power of BS's is based on a wireless communication bandwidth. When the number of antennas is configured as 4, Fig. 4(b) depicts the computation power of BS's with respect to bandwidths. Based on the results in Fig. 4(b), the computation power of BS increases with the increase of bandwidths. Moreover, the computation power of macro cell BS's is always larger than the computation power of small cell BS's when the bandwidth is increased. When the bandwidth is 400 MHz, i.e., adopting the millimeter wave technology, the computation power of macro cell BS is larger than 1000 W and the computation power of small cell BS is larger than 200 W. Based on results in Fig. 4, small cell BS's can save more computation power for BBU's than macro cell BSs in 5G mobile communication systems.

To evaluate the role of computation power in the BS, the computation power ratio is defined by the computation power over the total power at a BS. Fig. 5 illustrates the computation power ratio with respect to the number of antennas and bandwidths for small cell BS's and macro cell BS's. Fig. 5(a) shows the computation power with respect to the number of antennas. Based on the results in Fig. 5(a), the computation power ratio increases with the increased number of antennas. Moreover, the computation power ratio of small cell BS's is always larger than the computation power ratio of macro cell BS's. In addition, the computation power ratio of small cell BS's is obviously larger than 50\%. Fig. 5(b) depicts the computation power ratio with respect to bandwidths. Based on the results in Fig. 5(b), the computation power ratio increases with the increase of bandwidths. Moreover, the computation power of small cell BS's is always larger than the computation power of macro cell BS's. When millimeter wave technology is adopted, i.e., the bandwidth is larger than or equal to 20 MHz, and the computation power ratio of small cell BS's is obviously larger than 50\%.

\section{Future Challenges}
Based on the results in Fig. 4 and Fig. 5, the computation power will play a more important role than other power consumptions, including the transmission of power at 5G small cell BS's, no matter what the level of the absolute volume and the ratio for 5G small cell networks is. On the other hand, energy efficiency of 5G mobile communication systems is expected to improve 100 to 1000 times, compared with the energy efficiency of 4G mobile communication systems. However, most studies involving the energy efficiency of 5G cellular networks still focus on the transmission power optimization of BS's. To face the role of computation power in 5G small cell networks, some potential challenges are presented here.

The first challenge is the impact of 5G network architectures on the computation power in 5G small cell networks. Based on the results in Fig. 5, the importance of computation power is improved for energy efficiency optimization of 5G small cell networks. One obvious reason is that the transmission power is reduced in 5G small cell networks that adopt the massive MIMO and millimeter wave technologies. With cloud/fog computing and cache communications emerging for 5G networks, more and more data storage and computation will be performed at 5G small cell BS's. Therefore, it is possible to predict that computation power, no matter what the power consumption level of the absolute volume and the ratio will be shall further improve for 5G cellular networks. In this case, the energy efficiency optimization of 5G cellular networks will not only consider the transmission power. But also the power consumed for data computation and storage at BS's.

The second challenge is optimization of computation power at BS's with massive MIMO and millimeter wave transmission technologies. Existing studies usually fix the value of computation power at BS's. Moreover, the impact of 5G wireless transmission technologies, such as the massive MIMO and millimeter wave technologies on the computation power, is ignored at BS's. Based on the results in Fig. 4, the massive MIMO and millimeter wave technologies have a greater impact on the computation power of 5G small cell BS's. Considering the role of computation power at 5G small cell BS's, it is inadvisable to ignore the impact of 5G transmission technologies on the computation power of 5G small cell BS's. When massive MIMO and millimeter wave technologies are adopted by 5G small cell BS's, a large number of antennas and bandwidths can be scheduled for resource optimization in 5G small cell networks. How to schedule the number of antennas and bandwidths for the optimization of computation power at 5G small cell BS's.

The third challenge is the tradeoff between computation power and transmission power in 5G networks. Based on the analysis in Section II, the additional power of BS's depends on the computation and transmission powers of BS's. When the additional power of BS's is combined into the computation and transmission power of BS's, the energy efficiency of 5G networks can be calculated by the energy efficiency of computation and transmission powers at BS's. However, 5G transmission technologies have different effects on the energy efficiency of computation and transmission power of small cell BS's. In some specific scenarios, the effects on energy efficiency of computation and transmission powers are contradictory at 5G small cell BS's. Hence, the relationship between the computation and communication powers needs to be further investigated for 5G networks. Moreover, the tradeoff between computation and transmission power needs to be optimized for 5G small cell BS's.

To face the above challenges in the energy efficiency optimization of 5G small cell networks, some potential research directions are summarized to solve these issues:

\begin{itemize}
  \item
       The new energy efficiency model of 5G small cell networks considering computation and transmission power needs to be investigated. Moreover, the software-defined networks (SDN) could be used to trade off computation and transmission powers at 5G small cell BS's with cloud/fog computing functions.
  \item
      To improve the energy efficiency of 5G small cell BS's, joint optimization schemes and algorithms should be developed to save computation and transmission power at BBU's and RF chains together.
  \item
       Based on the simulation results in Fig. 4, lot of computation power of BBU's has to be changed into heat and more cooling systems need to be designed to support computation functions at BBU's. To save energy at BBU's, we should take the energy cycle into account and some potential technologies are expected to change the heat from BBU's into electrical energy based on the pyroelectric effect.
\end{itemize}

\section{Conclusions}
Until recently, the computation power of BS's was ignored or just fixed as a small constant in the energy efficiency evaluation of cellular networks. In this paper, the power consumption of BS's is analyzed for 5G small cell networks adopting massive MIMO and millimeter wave technologies. Considering the massive traffic in 5G small cell networks, the computation power of 5G small cell BS's is first estimated based on Landauer's principle. Moreover, simulation results show that the computation power of BS's increases as the number of antennas and bandwidths increases. Compared with transmission power, computation power will play a more important role in the energy efficiency optimization of 5G small cell networks. Therefore, we conclude that the energy efficiency optimization of 5G small cell networks should consider computation and transmission power together. How to converge computation and transmission technologies to optimize the energy efficiency of 5G networks is still an open issue. If this is accomplished, a different challenge would indeed emerge in the next round of the transmission and computation revolution.

\section*{Acknowledgment}
The authors would like to acknowledge the support from the NSFC Major International Joint Research Project (Grant No. 61210002), the Fundamental Research Funds for the Central Universities under the grant 2015XJGH011. This research is partially supported by the EU FP7-PEOPLE-IRSES, project acronym WiNDOW (grant no. 318992) and project acronym CROWN (grant no. 610524), China international Joint Research Center of Green Communications and Networking (No. 2015B01008).

\section*{Biographies}
XIAOHU GE  is currently a full professor with the School of Electronic Information and Communications at Huazhong University of Science and Technology (HUST), China, and an adjunct professor with the Faculty of Engineering and Information Technology at the University of Technology Sydney (UTS), Australia. He received his Ph.D. degree in communication and information engineering from HUST in 2003. His research interests include green communications, vehicular communications and wireless networks. He is the director of the China International Joint Research Center of Green Communications and Networking. He serves as
associate editors for IEEE Transaction on Greening Communications and Networking, IEEE Access.
\\

JING YANG received his B.E. degrees in communication engineering from HUST in 2014. He is currently pursuing the Ph.D. degree in the School of Electronic Information and Communications at HUST. His research interests mainly include computation power of wireless communication systems, green communication and energy efficiency of wireless cellular networks.
\\

HAMID GHARAVI (Life Fellow, IEEE) received his Ph.D. degree from Loughborough University, United Kingdom, in 1980. He joined the Visual Communication Research Department at AT\&T Bell Laboratories, Holmdel, New Jersey, in 1982. He was then transferred to Bell Communications Research (Bellcore), where he became a Distinguished Member of Research Staff. In 1993, he joined Loughborough University as a professor and chair of communication engineering. Since September 1998, he has been with the National Institute of Standards and Technology (NIST). His research interests include smart grid, wireless multimedia, mobile communications and wireless systems, mobile ad hoc networks, and visual communications. He served as a member of the Editorial Board of Proceedings of the IEEE from January 2003 to December 2008. From January 2010 to December 2013 he served as Editor-in-Chief of IEEE Transactions on CAS for Video Technology. He is currently serving as Editor-in-Chief of IEEE Wireless Communications.
\\

YANG SUN received his B.S. degrees with honor in electronic and information engineering from Huazhong university of science and technology and now is pursuing his master degree in electronic information and communications, Huazhong university of science and technology

\clearpage
\begin{figure}[!t]
\begin{center}
\includegraphics[width=6.5in]{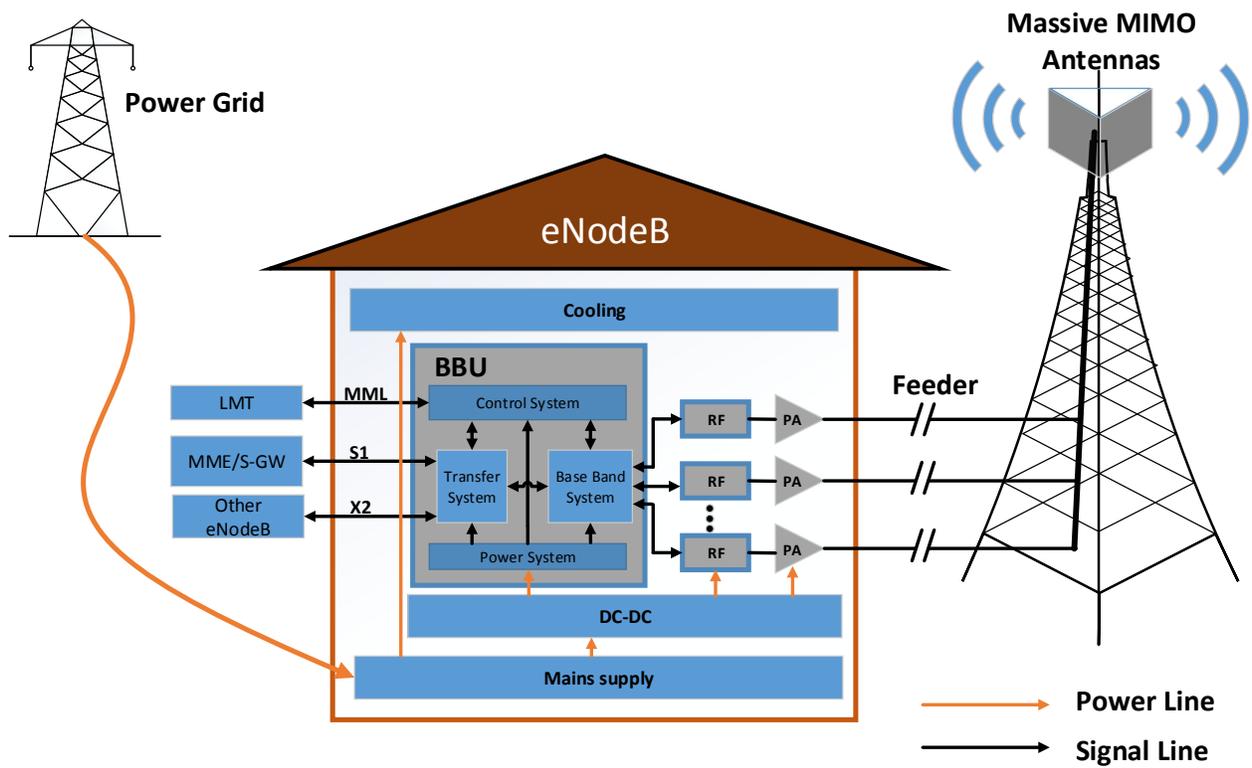}
\caption{Logistical architecture of eNodeB BS.}\label{Fig1}
\end{center}
\end{figure}

\clearpage
\begin{figure}[!t]
\begin{center}
\includegraphics[width=6.5in]{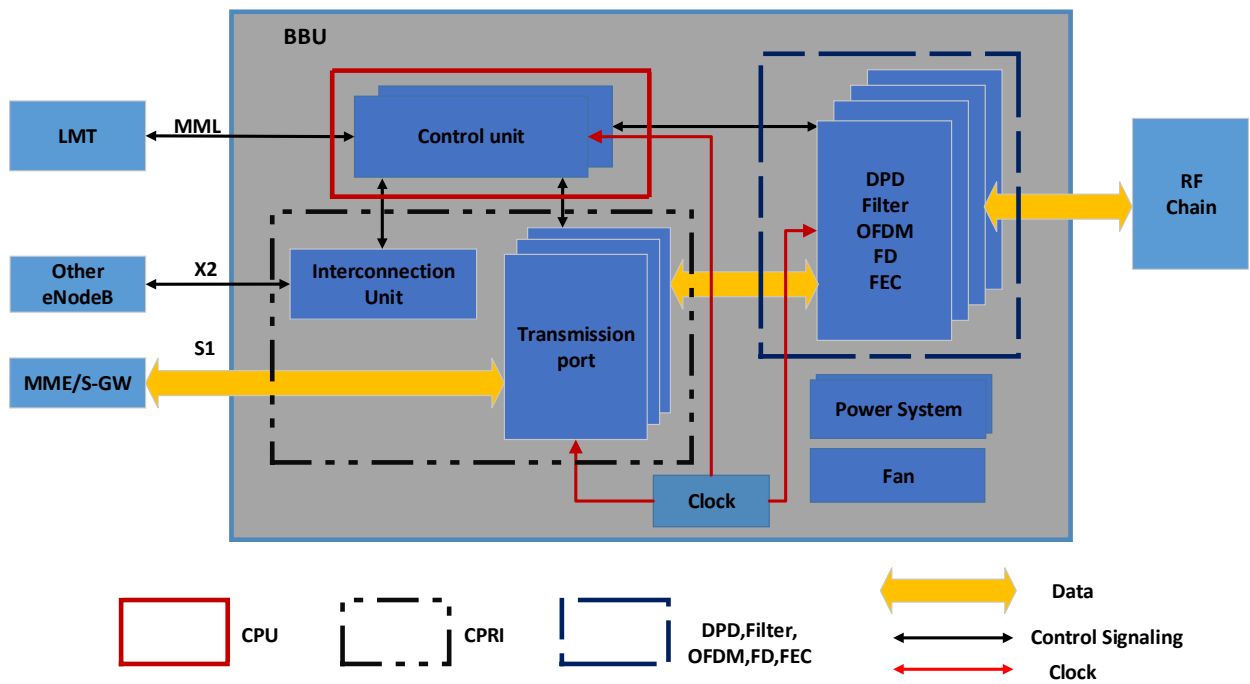}
\caption{Hardware architecture of BBU}\label{Fig2}
\end{center}
\end{figure}

\clearpage
\begin{figure}[!t]
\begin{center}
\includegraphics[width=6.5in]{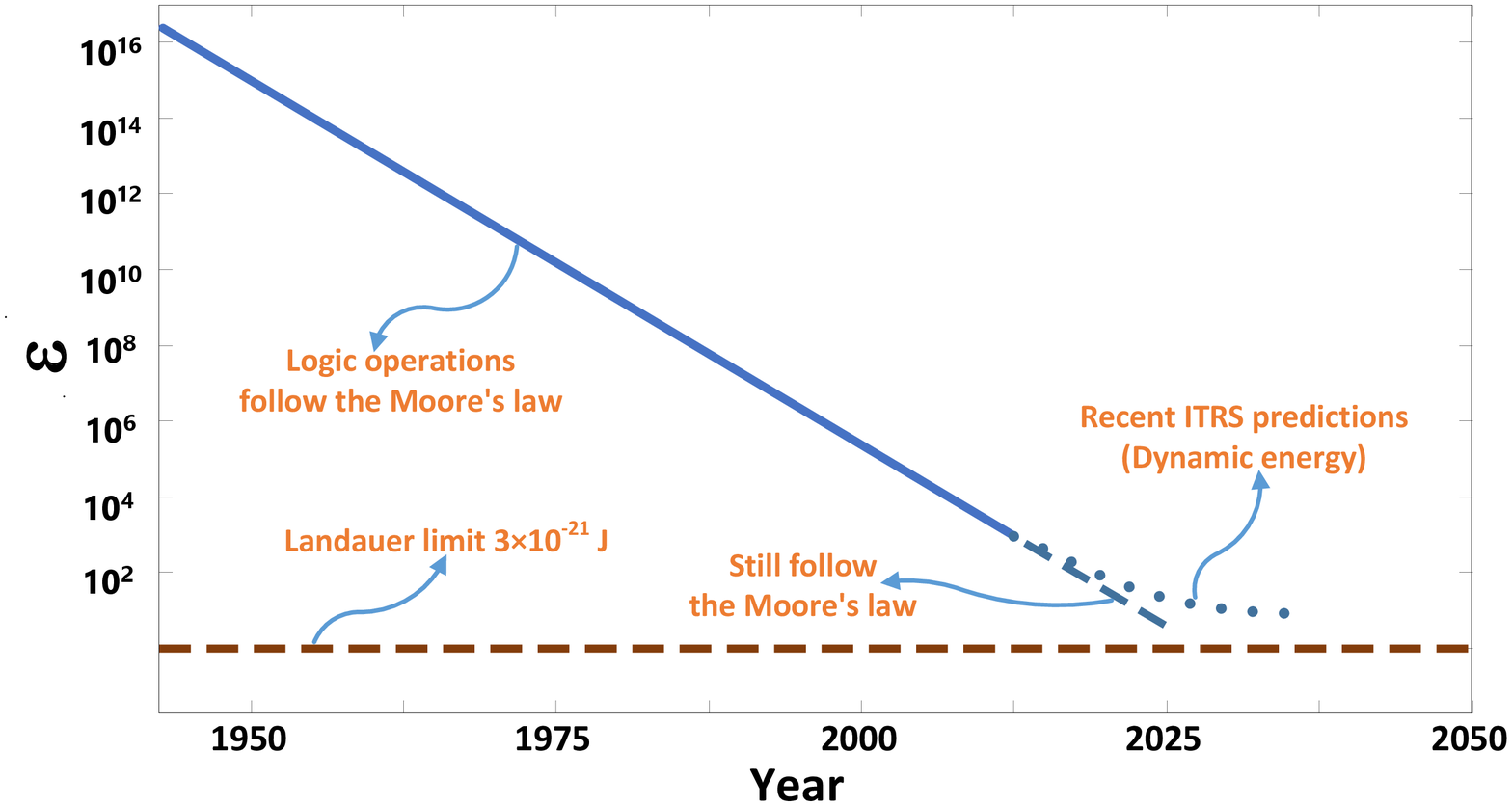}
\caption{The power coefficient with respect to the development of chip techniques}\label{Fig3}
\end{center}
\end{figure}

\clearpage
\begin{figure}[!t]
\begin{center}
\includegraphics[width=6.5in]{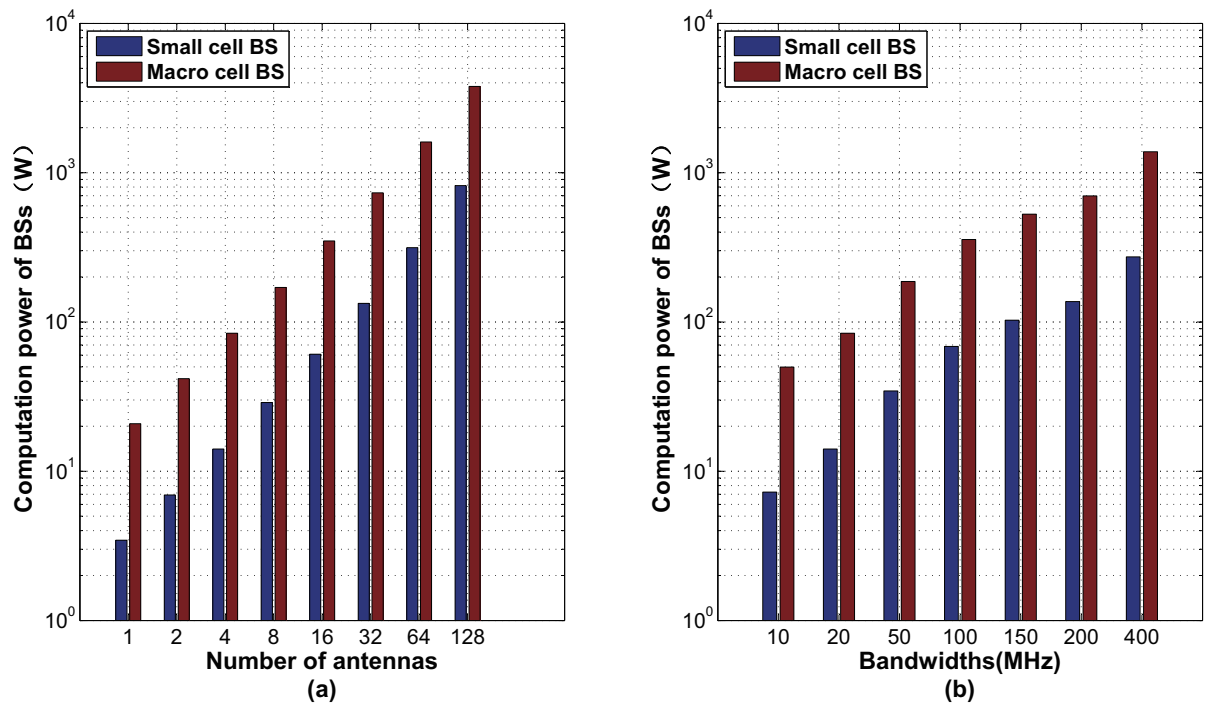}
\caption{Computation power of BSs with respect to the number of antennas and bandwidths}\label{Fig4}
\end{center}
\end{figure}

\clearpage
\begin{figure}[!t]
\begin{center}
\includegraphics[width=6.5in]{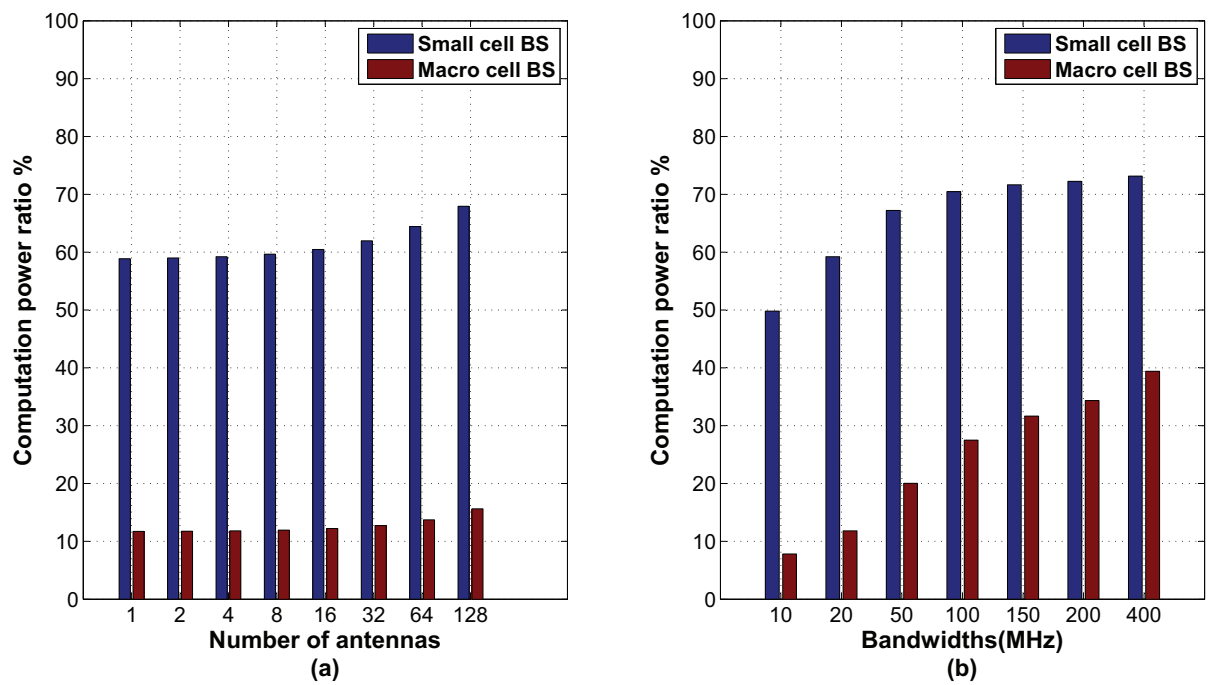}
\caption{Computation power ratio with respect to the number of antennas and bandwidths.}\label{Fig5}
\end{center}
\end{figure}

\clearpage
\begin{table}[t!]
\centering
\normalsize{

\caption{Configuration parameters of the BBU}\label{table2}

\begin{tabular}{p{2cm}<{\centering}|p{2cm}<{\centering}|p{2cm}<{\centering}|p{1cm}<{\centering}|p{1cm}<{\centering}|p{1cm}<{\centering}|p{1cm}<{\centering}|p{1cm}<{\centering}|p{1cm}<{\centering}}

\hline
\textbf{BBU }  &\textbf{GOPS of}  &\textbf{GOPS of}  &\textbf{${S_{BW}}$}  &\textbf{${S_M}$}  &\textbf{${S_R}$}  &\textbf{${S_{Ant}}$}  &\textbf{${S_{dt}}$}  &\textbf{${S_{df}}$} \\

\textbf{parameters}  &\textbf{Macro cell}  &\textbf{Small cell}  &  &  &  &  &  & \\

\hline
DPD	&160  &0  &1  &0  &0  &1  &1  &0\\
\hline
Filter	&400  &250  &1  &0  &0  &1  &1  &0\\
\hline
CPRI/SERDES	&720  &0  &1  &1  &1  &1  &1  &1\\
\hline
OFDM  &160  &120  &1  &0  &0  &1  &1  &0\\
\hline
FD  &90  &50  &1  &0  &0  &1  &1  &1\\
(liner)  &  &  &  &  &  &  &  &\\
\hline
FD	&30  &15  &1  &0  &0  &2  &1  &1\\
(non-liner)	&  &  &  &  &  &  &  &\\
\hline
FEC 	&140  &130  &1  &1  &1  &1  &1  &1\\
\hline
CPU	&400  &40  &0  &0  &0  &1  &0  &0\\
\hline

\end{tabular}
}
\vspace{0.8cm}\\
\end{table}

\end{document}